\documentclass[prc,showpacs,showkeys,superscriptaddress,nofootinbib,twocolumn,floatfix]{revtex4-1}
\usepackage{amsfonts}
\usepackage{float}
\usepackage{placeins}
\usepackage{graphicx}
\usepackage[hidelinks]{hyperref}
\usepackage{color,amsmath,amssymb,bm}
\usepackage{tensor}

\usepackage[normalem]{ulem}  

\renewcommand{\sout}{\bgroup \color{red} \ULdepth=-.5ex \ULset}

\def\blfootnote{\xdef\@thefnmark{}\@footnotetext}

\newcommand{\beq}{\begin{equation}}
\newcommand{\eeq}{\end{equation}}
\newcommand{\bea}{\begin{eqnarray}}
\newcommand{\eea}{\end{eqnarray}}

\begin{document}


\title{Modification of  $Z^0$ leptonic invariant mass in ultrarelativistic heavy ion collisions as a measure of the electromagnetic field}

\author{Yifeng Sun}
\email{sunyfphy@lns.infn.it}
\affiliation{Department of Physics and Astronomy, University of Catania, Via S. Sofia 64, 1-95125 Catania, Italy}
\affiliation{Laboratori Nazionali del Sud, INFN-LNS, Via S. Sofia 62, I-95123 Catania, Italy}

\author{Vincenzo Greco}
\affiliation{Department of Physics and Astronomy, University of Catania, Via S. Sofia 64, 1-95125 Catania, Italy}
\affiliation{Laboratori Nazionali del Sud, INFN-LNS, Via S. Sofia 62, I-95123 Catania, Italy}

\author{Xin-Nian Wang}
\affiliation{Nuclear Science Division, Lawrence Berkeley National Laboratory, Berkeley, CA 94720, USA}

\date{\today}

\begin{abstract}
An extraordinary strong magnetic field, $eB_0 \approx 10^{18}$ Gauss, is expected to be generated in non-central ultrarelativistic heavy ion collisions 
and it is envisaged to induce several effects on hot QCD matter including the possibility of local parity and 
local charge conjugation and parity  symmetry violations. 
A direct signature of such e.m. fields and a first quantitative measurement of its strength and lifetime are still missing.
We point out that both the mean value of leptonic invariant mass of $Z^0$ boson, reconstructed by its decaying lepton pairs, and the relative width
are modified in relativistic heavy  ion collisions due to the presence of strong  initial  e.m. fields. 
We propose a measurement of the leptonic invariant mass of $Z^0$ as  a novel probe of the strength of the $B_y$. Both shifts  could be up to about few hundred MeV and are found to depend on the integral of $B_y$ over the time duration quadratically (approximate).
Hence it provides a novel and clear probe of electromagnetic fields, which can be tested experimentally.
\end{abstract}

\maketitle

\section{Introduction}

The ultrarelativistic heavy ion collisions (uRHICs) experiments conducted at both the BNL Relativistic Heavy Ion Collider (RHIC)~\cite{Adams:2005dq,Adcox:2004mh} and the CERN Large Hadron Collider (LHC)~\cite{Aamodt:2008zz} have created a new state of matter with deconfined quarks and gluons, the quark-gluon plasma (QGP). The QGP is found to be the most perfect fluid created in nature~\cite{Kovtun:2004de,Romatschke:2007mq,Gale:2013da}. Heavy ion collisions also provide the possibility to probe the local parity (P) as well as  charge conjugation and parity (CP) symmetry violation processes in Quantum chromodynamics (QCD) that may be generated by the metastable local domains of gluon fields with a non-zero winding number~\cite{PhysRevD.8.1226,PhysRevD.9.2291,PhysRevLett.81.512}. The most promising way to probe the P and CP violations in QCD is to measure the chiral magnetic effect (CME)~\cite{Kharzeev:2007jp,Fukushima:2008xe,Kharzeev:2009fn,Jiang:2016wve,Shi:2017cpu,Sun:2018idn,PhysRevLett.125.242301}, where a strong magnetic field with a long lifetime is required in order to generate a signal large enough. 

A huge electromagnetic field can be generated in non-central ultrarelativistic heavy ion collisions. However, there are a lot of 
inherent uncertainties in the calculation of the time evolution of the magnetic field in heavy ion collisions due to the uncertainty of the electrical conductivity of QGP~\cite{Ding:2010ga,Amato:2013naa,Brandt:2012jc}, the poor knowledge of the properties of the initial non-equilibrium stage as well as  the complexity of numerically solving magnetohydrodynamics (MHD). This inspired the search for a direct probe of the strong e.m. fields by measuring the directed flow $v_1=\langle p_x/p_T \rangle$ splitting between positively and negatively charged hadrons ~\cite{Gursoy:2014aka,Gursoy:2018yai}, 
especially heavy meson pairs ($D^0,\overline{D}^0$) ~\cite{Das:2016cwd,Chatterjee:2018lsx,Sun:2020wkg} 
or the leptons decayed from $Z^0$ boson ~\cite{Sun:2020wkg}. 
Here we propose a new probe of electromagnetic fields via the leptonic invariant mass distribution of $Z^0$ boson reconstructed from its decaying lepton pairs, whose final momenta should be affected by the presence of e.m. fields. 
It should be significantly easier to measure the invariant mass distribution of $Z^0$ boson than the $v_1(p_T)$ splitting between its decaying leptons of opposite charge. Hence the measurement of the invariant mass distribution of $Z^0$ would open up a more accessible experimental probe
that as we discuss in this Letter can be directly linked to the time integral of the magnetic field.

The paper is organized as follows: In Sec. II we discuss the parametrization of e.m. fields and a description of the coordinate and momentum 
distributions of both $Z^0$ boson and its decaying lepton pairs. In Sec. III we present several numerical results including the invariant mass distribution of $Z^0$ reconstructed by its decaying lepton pairs in the presence of e.m. fields and the dependence of the shifts of both the invariant mass of $Z^0$ and its width on the configuration of e.m. fields. Summary and conclusions are given in Sec. IV.

\section{electromagnetic fields and Leptons from $Z^0$}
To study the effect of e.m. fields on the $Z^0$ invariant mass reconstructed by its decaying lepton pairs, we adopt a general parametrization of the configurations of e.m. fields used in several studies~\cite{Roy:2017yvg,Shi:2017cpu,Sun:2018idn,Burnier:2011bf,Jiang:2016wve}: 
\begin{eqnarray}
&&eB_y(x,y,\tau)=-B(\tau)\rho_B(x,y)
\\&&\rho_B(x,y)=\rm{exp}[-\frac{x^2}{2\sigma_x^2}-\frac{y^2}{2\sigma_y^2}]
\\&&B(\tau)=eB_0/(1+(\tau/\tau_B)^a),
\end{eqnarray}
 where $B_0$, $\sigma_x$ and $\sigma_y$ are usually given by the estimates of e.m. fields in the vacuum in AA collisions at $t=0$~\cite{Deng:2012pc}. The above gives the transverse coordinate dependence and time evolution of $B_y$.
 The electric field $eE_x$ is then determined by solving the Faraday's Law $\boldsymbol{\nabla}\times\mathbf{E}=-\partial \mathbf{B}/\partial t$:
\begin{eqnarray}
&&eE_x(t,x,y,\eta_S)=\rho_B(x,y)\int_0^{\eta_S}d\chi B^{'} \left(\frac{t}{\rm{cosh}\chi}\right) \frac{t}{\rm{cosh}\chi}
\label{eEx}
\end{eqnarray}
where the invariant time $\tau$ and space-time rapidity $\eta_S$ are related to $t$ and $z$ by $\tau=\sqrt{t^2-z^2}$ and $\eta_S=\frac{1}{2} \ln(\frac{t+z}{t-z})$. We note that the above configurations of e.m. fields may not apply to space with a large magnitude of $\eta_S$ and transverse coordinate $\rho=\sqrt{x^2+y^2}$, where one should solve the full Maxwell equations with complex boundary conditions. However, we can safely adopt the above configurations of e.m. fields at small magnitude of $\eta_S$ and $\rho$ considering 
initial transverse coordinates of particles are mostly centered in the overlapping region making  the detailed behavior
of the e.m. fields at large $\rho$ irrelevant.  

We will focus on 5.02 TeV PbPb collisions at 20-30\% centrality  for the numerical calculations, which corresponds to impact parameter $b=7.5$ fm. However, the conclusions should be general based on our physical arguments. The parameters in this colliding system are found to be $eB_0=73\, m_{\pi}^2$, $\sigma_x=3$ fm and  $\sigma_y=4$ fm~\cite{Sun:2020wkg}, where $eB_0$ is the maximum initial value estimated in the vacuum.

The distribution of $Z^0$ boson in the transverse plane  is given by the binary nucleon-nucleon collisions of colliding nuclei, while in the longitudinal axis $z=\tau_{Z^0}\sinh y_z$ and $t=\tau_{Z^0}\cosh y_z$ with $\tau_{Z^0}=\hbar/m_{Z^0}=0.0022$ fm$/c$, where $y_z$ is the rapidity of $Z^0$. The momentum distribution of $Z^0$  in 5.02 TeV Pb+Pb collisions is given by fitting the experimental measurements~\cite{Chatrchyan:2014csa,Khachatryan:2015pzs} with:
\begin{eqnarray}
&&\frac{dN}{d^2p_Tdy_z}=f(\mathbf{p_T},y_z)\propto 10^{-ap_T^{n}}e^{-\frac{y_z^2}{2 \Delta_l^2}},
\end{eqnarray}
where $a=0.6896$, $n=0.4283$ and $\Delta_l=3.034$ are found to give quite a good description of $p_T$  and $y_z$  
dependence of $Z^0$ boson, as in reported in ~\cite{Sun:2020wkg}.

We use the Monte Carlo method to generate $Z^0$ boson, whose invariant mass distribution  is given by a Breit-Wigner distribution:
\begin{eqnarray}
&&\rho(M)=\frac{1}{\pi}\frac{\Gamma/2}{(M-M_0)^2+\Gamma^2/4},
\label{BW}
\end{eqnarray}
with $M_0=91.1876$ GeV and $\Gamma=2.4952$ GeV~\cite{PhysRevD.98.030001}. Finally the spacetime coordinate of produced lepton pairs is given by their mother $Z^0$ that moves in a straight line, with the decay time following  a distribution $\rho(\Delta t)\propto e^{-\frac{\Gamma\Delta t}{\gamma_v}}$, with $\gamma_v$ being the Lorentz contraction factor. After the interaction with e.m. fields, these lepton pairs are used to reconstruct the invariant mass of $Z^0$ boson.

\section{numerical results}
\subsection{The effect of lepton-quark scattering}
Before discussing the effect of external e.m. fields on the $Z^0$ boson invariant mass distribution, it should be noted that this distribution can also be modified due to the lepton-quark scattering in QGP~\cite{PhysRevLett.122.132301}. To consider this effect, 
we employ the standard Langevin equations:
\begin{eqnarray}
&&d x_i = \frac{p_i}{E}dt,
\\&&{d p_i} = -\gamma p_idt+\xi_{i}\sqrt{2D_pdt},
\label{Langevin}
\end{eqnarray}
where the momentum diffusion coefficient $D_p$ is related to the drag coefficient $\gamma$, energy of leptons $E$, and the local temperature $T$ by $D_p=\gamma ET$, and $\xi_i$ is a real number randomly sampled from a normal distribution with $\langle\xi_i\rangle=0$ and $\langle \xi_i\xi_j\rangle=\delta_{ij}$. $D_p$ is related to the transverse momentum broadening rate $\hat{q}$ due to elastic collisions between leptons and the medium quarks by $D_p=\hat{q}/4$~\cite{Liu:2021dpm}. Since the small angle scattering cross section for lepton-quark scattering is:
\begin{eqnarray}
&&\frac{d\sigma}{dq_{\perp}^2}\approx e_q^2\frac{2\pi\alpha_e^2}{q_{\perp}^4},
\end{eqnarray}
$\hat{q}$ will be:
\begin{eqnarray}
\hat{q}&=&\sum_q\int_{\mu^2}^{s^*/4} dq_{\perp}^2 \rho_q e_q^2 \frac{d\sigma}{dq_{\perp}^2} q_{\perp}^2\nonumber
\\&=&\frac{12\zeta(3)}{\pi}\alpha_e^2T^3 \ln{\frac{s^*}{4\mu^2}},
\end{eqnarray}
where $\alpha_e$ is the fine structure constant in  QED, $\zeta(3)\approx1.202$, $\rho_q$ is the number density of quarks of each flavor, $s^*\approx 5.6ET$ is the average center of mass energy of lepton-quark scattering through one photon exchange, and $\mu^2=\frac{1}{2}(3+N_c\sum_q e_q^2)e^2T^2=10\pi\alpha_eT^2$ is the Debye screening mass for the exchange photon from quark and lepton loops .

To quantitatively characterize the effect of lepton-quark scattering or the e.m. fields on the invariant mass of $Z^0$ boson, we define two quantities:
\begin{eqnarray}
&&\Delta \langle M \rangle=\langle M_f \rangle-\langle M_i \rangle
\\&&\Delta \sigma=\sigma_f-\sigma_i\nonumber
\\&&=\sqrt{\frac{\sum (M_f-\langle M_f\rangle)^2}{N-1}}-\sqrt{\frac{\sum (M_i-\langle M_i\rangle)^2}{N-1}},
\end{eqnarray}
where $f$ and $i$ stand for the invariant mass of $Z^0$ reconstructed by lepton pairs in vacuum and with the effect of lepton-quark scattering or the e.m. fields, and $N$ is the number of $Z^0$ boson used in calculation.

After the evolution of leptons in QGP due to lepton-quark scattering
described by Eq. (\ref{Langevin}), the results show $\Delta \langle M \rangle=-1.9$ MeV and $\Delta\sigma\le0.2$ MeV, which is a small number compared to the experimental uncertainty on $M_0$ and $\Gamma$ and the modification due to e.m. fields as we will show below. We thus do not include this in the following discussions of the effects of e.m. fields. However, it should be noted that this effect is stronger in more central collision because the lifetime is longer and the temperature of QGP is higher, while the effect of e.m. fields should be smaller because the magnetic field decreases in more central collisions.

\subsection{Relating the leptonic invariant mass and width of $Z^0$ to e.m. fields strength}
\begin{figure}[h]
\centering
\includegraphics[width=1\linewidth]{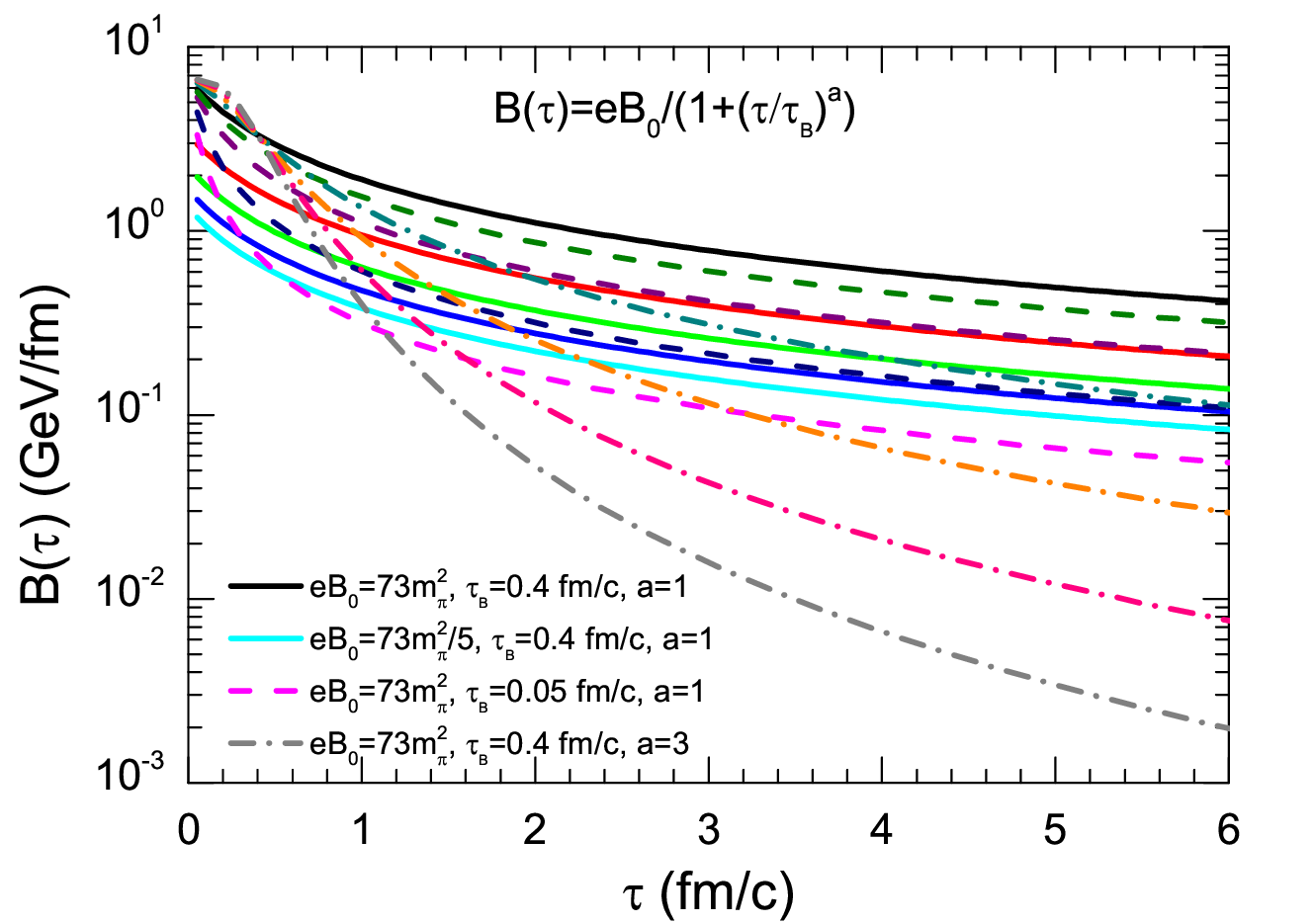}
\caption{(Color online)  Time evolution of $B(\tau)$ with different sets of $eB_0$, $\tau_B$ and $a$, where $eB_0$, $\tau_B$ and $a$ increase from $73m_{\pi}^2/5$ to $73 m_{\pi}^2$,  from 0.05 fm$/c$ to 0.4 fm$/c$ and  from 1 to 3, respectively.}
\label{fig:Bt}
\end{figure}

In Fig. \ref{fig:Bt} we show the time evolution of $B(\tau)$ with different sets of $eB_0$, $\tau_B$ and $a$, 
where one can see that a wide range of e.m. fields is explored.
We do not show the time evolution of $E_x$, calculated by Eq. (\ref{eEx}), which also varies 
according to different sets of $eB_0$, $\tau_B$ and $a$.

\begin{figure}[h]
\centering
\includegraphics[width=1\linewidth]{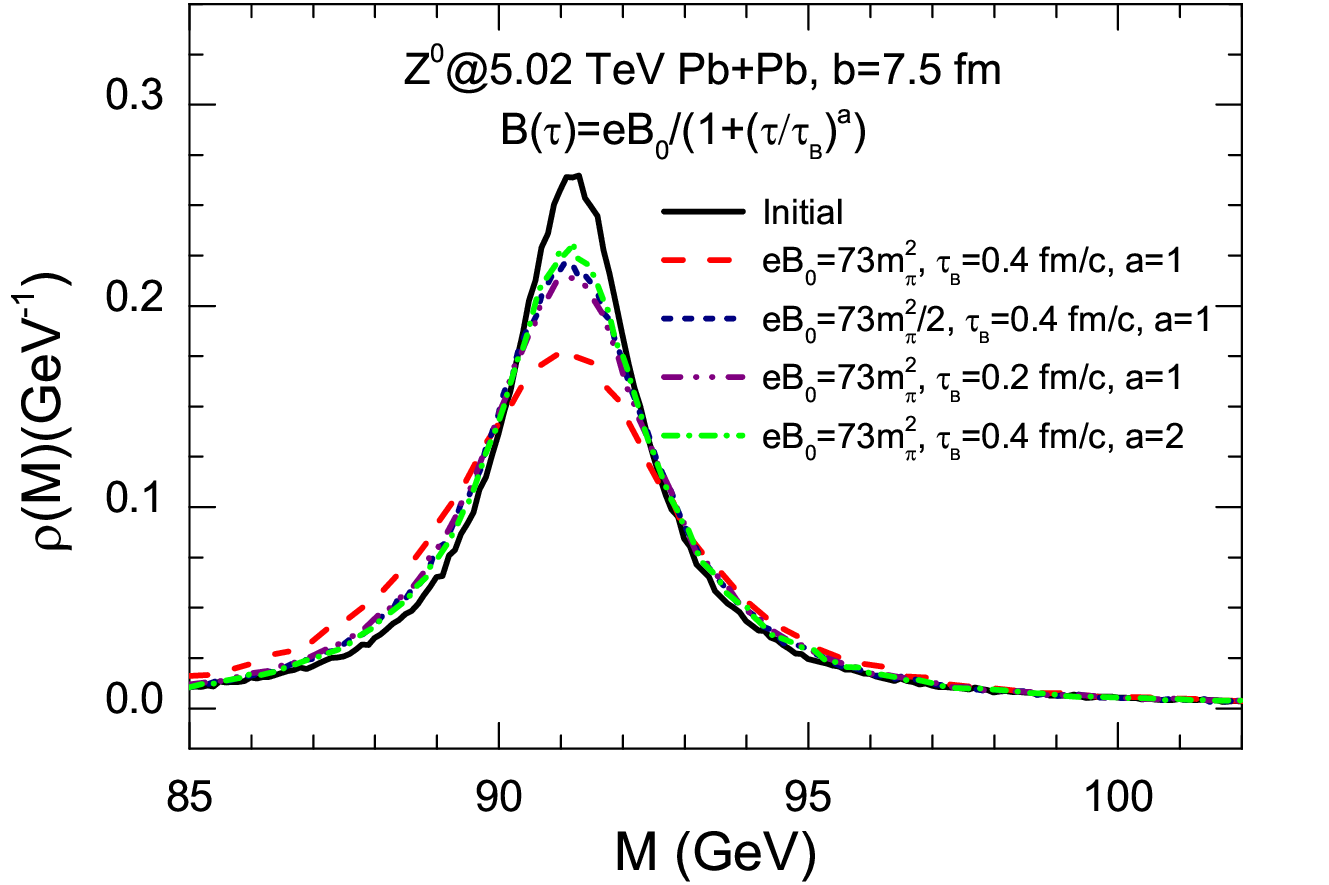}
\caption{(Color online)  The invariant mass distribution of $Z^0$ boson at midrapidity $|y_z|\le0.5$ reconstructed by lepton pairs after interacting with e.m. fields.}
\label{fig:rhom}
\end{figure}

The results of the invariant mass distribution of $Z^0$ are shown in Fig. \ref{fig:rhom}, where the solid black line shows the initial distribution of $Z^0$ invariant mass in vacuum, which has a Breit-Wigner form as in Eq. (\ref{BW}). The red line shows the distribution of the $Z^0$ invariant mass reconstructed from lepton pairs after interacting with e.m. fields with $eB_0=73 m_{\pi}^2, \tau_B=0.4$ fm$/c$ and $a=1$. 
This set of parameters is found to reproduce the directed flow splitting between $D^0$ and $\overline{D}^0$ 
with $d\Delta v_1/d\eta=0.49\pm0.17 (stat.) \pm 0.06 (syst.)$ 
as measured by the  ALICE experiment~\cite{Acharya:2019ijj}. It is seen by the red dashed line that such an e.m. field 
would strongly increase the width $\sigma_{Z^0}$ of the distribution of $Z^0$ invariant mass by about 300 MeV 
and decrease the mean value  $\langle M_{Z^0} \rangle$ by about 250 MeV. We have varied $eB_0$, $\tau_B$ and $a$ by a factor of two, respectively. The results are shown by the navy, purple and green lines in Fig.\ref{fig:rhom}, where it is seen that the width increases as well but not as much as the red line. 
The large uncertainty of ALICE measurements on the $v_1$ splitting of $D^0$ 
does not allow a determination of the e.m. field. Currently it is still
to be clarified whether $\Delta v_1^{D}$ is determined only by the e.m. fields \cite{Sun:2020wkg}.
Therefore to have a comprehensive study of the effect of e.m. fields on the invariant mass of $Z^0$ reconstructed by lepton pairs, we vary $eB_0$, $\tau_B$ and $a$ in $B(\tau)$ to find some general pattern relating $\Delta \langle M_{Z^0} \rangle$ and $\Delta\sigma_{Z^0}$
to the strength and time dependence of the magnetic field. We 
vary $eB_0$ by a factor of 5 and the life time $\tau_B$ by a factor of 8 and the power law parameter $a$ by a factor of 3 respectively, 
while keeping other parameters unchanged.


\begin{figure}[h]
\centering
\includegraphics[width=1\linewidth]{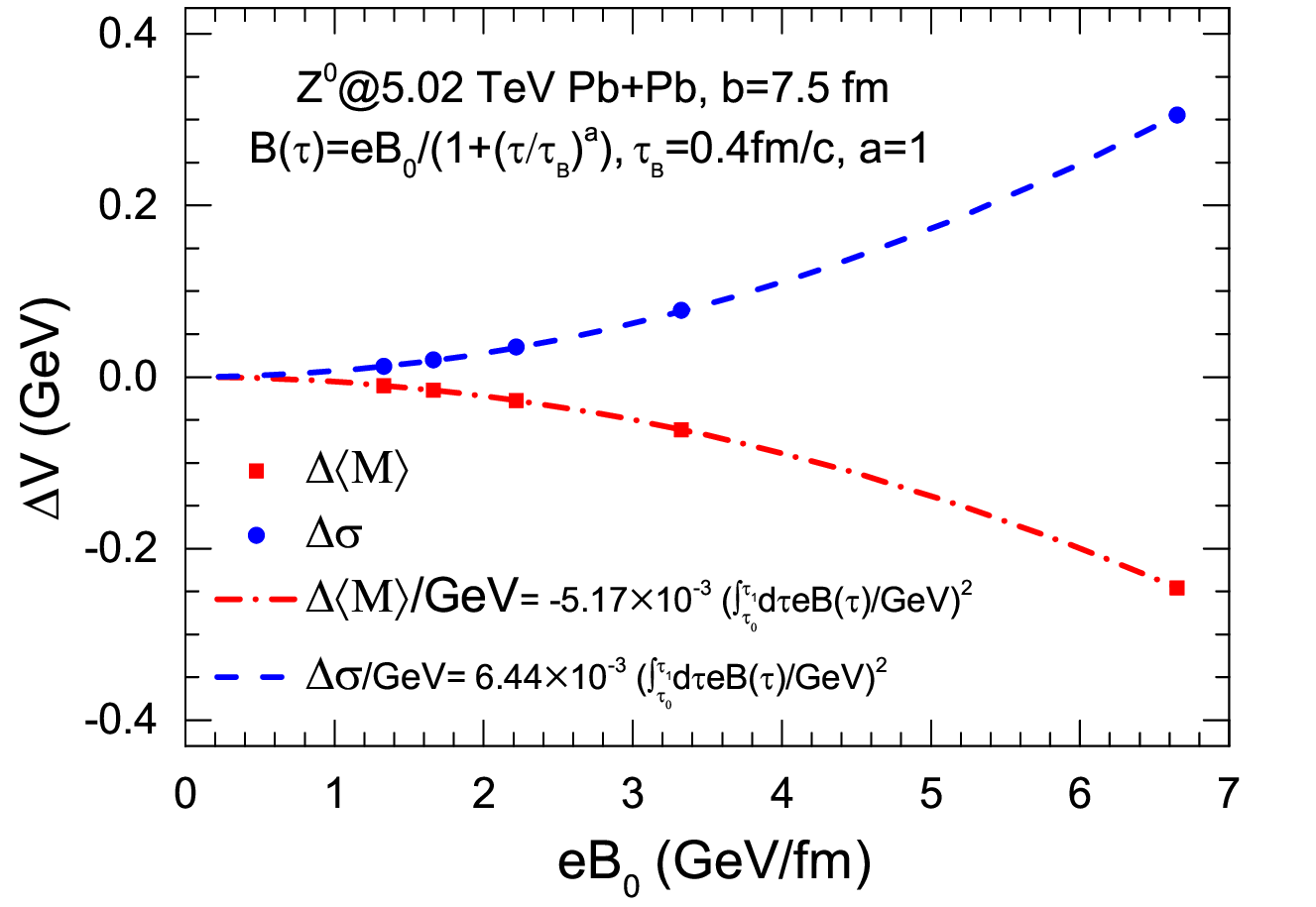}
\caption{(Color online)  The $eB_0$ dependence of $p_T$ integrated $\Delta \langle M \rangle$ and $\Delta \sigma$ of midrapidity ($|y_z|\le 0.5$) $Z^0$ boson induced by e.m. fields.}
\label{fig:changeeB0}
\end{figure}

In Fig. \ref{fig:changeeB0}, we show  how $p_T$ integrated $\Delta \langle M \rangle$ and $\Delta \sigma$ of   $Z^0$ boson in midrapidity ($|y_z|\le 0.5$) changes with $eB_0$, keeping $\tau_B=0.4 \,\rm fm/c$ and $a=1$.
It is seen by the red squares and blue circles  that $\Delta \langle M \rangle$ changes from -9.9 MeV to -246 MeV, and $\Delta \sigma$ from 12.6 MeV to 305 MeV, with $eB_0$ increasing from $73 m_{\pi}^2/5$ to  $73 m_{\pi}^2$. 

Because the invariant mass of $Z^0$ boson is symmetric with charge conjugation, $\Delta \langle M \rangle$ should be proportional to $(eB_0)^2$ in the leading order. More specifically, supposing one $Z^0$ boson at rest with mass $M$ decays into lepton pairs whose momenta $\boldsymbol{p}$ and $-\boldsymbol{p}$ change by $\Delta \boldsymbol{p}_1$ and $\Delta \boldsymbol{p}_2$ due to e.m. fields, then the invariant mass will change by:
\begin{eqnarray}
&&\Delta M=M_f-M=\nonumber
\\&&\sqrt{(E( \boldsymbol{p}+\Delta \boldsymbol{p}_1)+E( -\boldsymbol{p}+\Delta \boldsymbol{p}_2))^2-(\Delta \boldsymbol{p}_1+\Delta \boldsymbol{p}_2)^2}\nonumber
\\&&-M\approx\frac{(\Delta \boldsymbol{p}_1-\Delta \boldsymbol{p}_2 )^2+4\boldsymbol{p}\cdot(\Delta \boldsymbol{p}_1-\Delta \boldsymbol{p}_2 )}{2M},
\label{dm}
\end{eqnarray}
with $E(\boldsymbol{p})=\sqrt{m_l^2+\boldsymbol{p}^2}$.
The negative value of $\Delta \langle M \rangle$ implies thus $\langle \boldsymbol{p}\cdot(\Delta \boldsymbol{p}_1-\Delta \boldsymbol{p}_2 )\rangle<0$,
noting that in general $\Delta\boldsymbol{ p}_1 \neq \Delta\boldsymbol{ p}_2$.


The time integral $\int_{\tau_0}^{\tau_1}d\tau eB(\tau)$ should be a good quantity to qualify the effect of e.m. fields, where $\tau_0$ is the production time of lepton pairs that is about 0.08 fm$/c$ and $\tau_1$ is the effective time  when charged particles escape e.m. fields, which is about 6-8 fm$/c$ in semi-peripheral collisions. 
We found that both $\Delta \langle M \rangle$  and $\Delta \sigma$ can be simply fitted as  $k (\int_{\tau_0}^{\tau_1}d\tau eB(\tau))^2$ 
with $k_M=-5.17\times 10^{-3}$ for the mass, shown as the red dash-dotted line in Fig.~\ref{fig:changeeB0},  
and for the width $k_\sigma=6.44\times 10^{-3}$, as the blue dashed line.

\begin{figure}[h]
\centering
\includegraphics[width=1\linewidth]{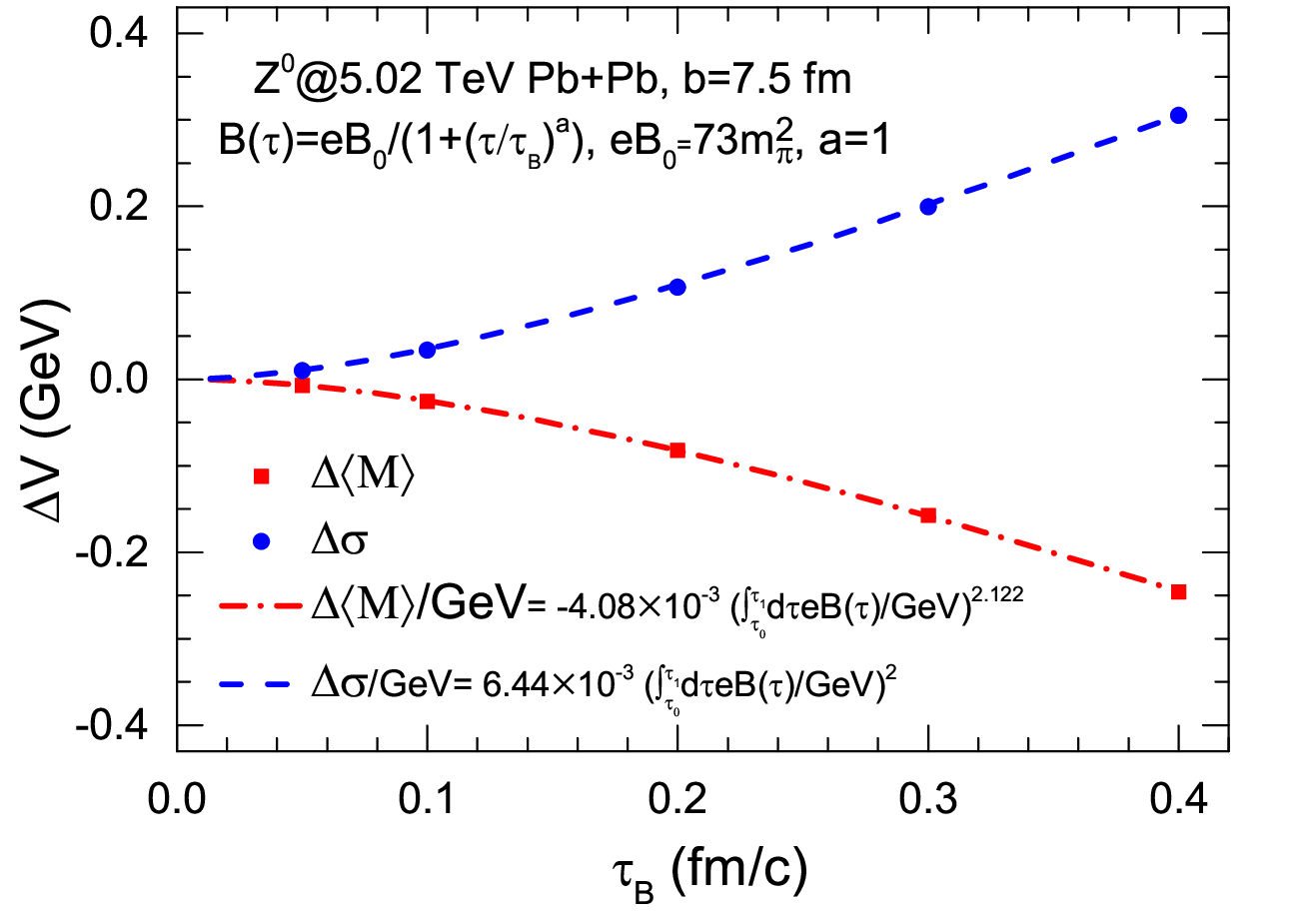}
\caption{(Color online)  The $\tau_B$ dependence of $p_T$ integrated $\Delta \langle M \rangle$ and $\Delta \sigma$ of midrapidity ($|y_z|\le 0.5$) $Z^0$ boson induced by e.m. fields.}
\label{fig:changetauB}
\end{figure}

In Fig. \ref{fig:changetauB}, we extend the study to $\tau_B$ dependence of the $p_T$ integrated $\Delta \langle M \rangle$ and $\Delta \sigma$ of $Z^0$ boson in midrapidity, varying $\tau_B$ by a factor of 8 from 0.05 fm/c to 0.4 fm/c. 
Because it is not trivial that the effects of $E_x$ and $B_y$  change by the same factor 
if one changes $\tau_B$, we fit $\Delta \langle M \rangle$ and $\Delta \sigma$ of $Z^0$ boson 
by $k(\int_{\tau_0}^{\tau_1}d\tau eB(\tau))^n$, where $n$ is not fixed to 2. However, we
found $\Delta \langle M \rangle$ is still nearly proportional to the square of the integral  with the fit parameter  $n_M=2.122$
and $k_M=-4.08\times 10^{-3}$, as shown by the red dash-dotted line. 
Instead $\Delta \sigma$ as a function of $\tau_B$ has the fitting parameters $k_\sigma$ and $n_\sigma=2$ that are the same as the previous case
studied as a function of $eB_0$, see the blue dashed line. 
According to the red squares and blue circles, $\Delta \langle M \rangle$ changes from -7.46 MeV to -246 MeV, 
and $\Delta \sigma$ from 10.2 MeV to 305 MeV, with $\tau_B$ increasing from 0.05 fm$/c$ to  0.4 fm$/c$.  

\begin{figure}[h]
\centering
\includegraphics[width=1\linewidth]{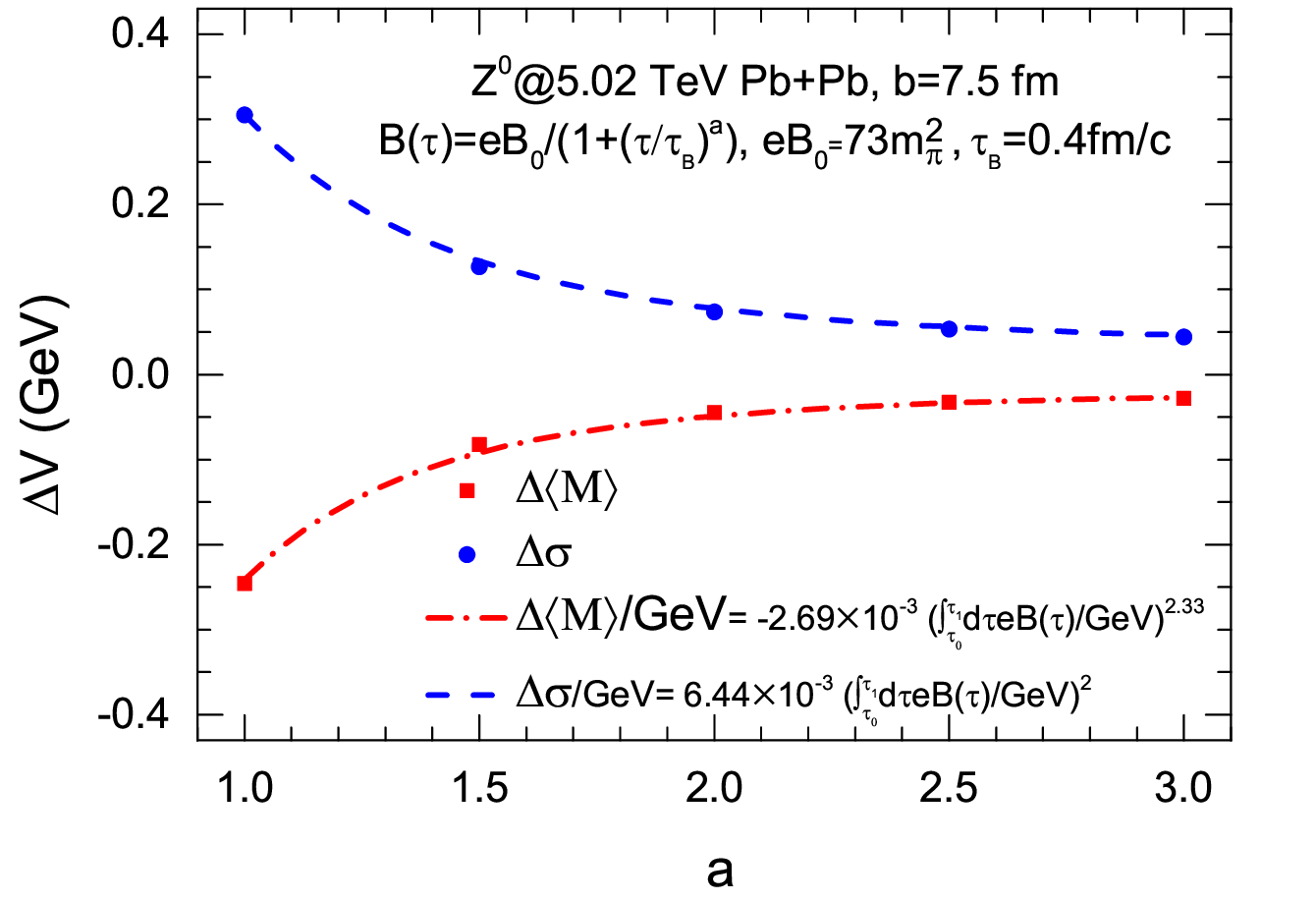}
\caption{(Color online)  The $a$ dependence of $p_T$ integrated $\Delta \langle M \rangle$ and $\Delta \sigma$ of midrapidity ($|y_z|\le 0.5$) $Z^0$ boson induced by e.m. fields.}
\label{fig:changepower}
\end{figure}

Finally, we vary the power law decay parameter $a$ in $B(\tau)$ by a factor of 3 which implies a very large change in the time dependence.
The red squares and blue circles in Fig. \ref{fig:changepower} show that $\Delta \langle M \rangle$ changes from -246 MeV to -27.9 MeV, and $\Delta \sigma$ from 305 MeV to 44.1 MeV, with $a$ increasing from 1 to 3. Moreover, as shown by the red dash-dotted and blue dashed lines in Fig. \ref{fig:changepower}, $\Delta \langle M \rangle$ is fitted well with $k_M=-2.69\times 10^{-3}$  and $n_M=2.33$ which
however stays still quite close to 2, while
the parameters $k_\sigma$ and $n_\sigma$ used in fitting $\Delta \sigma$ as a function of $a$ are found
again and quite remarkably to be the same as the other two cases, hence the quadratic relation remain a solid general relation.

In principle one may think to correlate the shifts of the mass $\Delta \langle M \rangle$ and $\Delta \sigma$ of $Z^0$ 
with the splitting in the directed flow $d\Delta v_1^{l}/dy_z|_{y_z=0}$
of the leptons of opposite charge as has been studied in \cite{Sun:2020wkg}, however it has to be noticed that the latter depends only on the
$d\Delta p_x/dy_z$, while the invariant mass distribution depends on all the vector components of the shift, 
 according to Eq. (\ref{dm}).
We have carried on a first study that finds a significant correlation, but only when $d\Delta v_1^{l}/dy_z|_{y_z=0}>0.05$, i.e.
the $\Delta p_x$ remains dominant,
but the correlations weakens
when it has smaller positive and negative values. A more detailed analysis about this aspect will be published later.

We also notice  that $\langle\Delta \boldsymbol{p}_1-\Delta \boldsymbol{p}_2\rangle$ should be zero due to P symmetry, but
if one looks at the $y_z$ dependence of $\langle\Delta \boldsymbol{p}_1-\Delta \boldsymbol{p}_2 \rangle$, 
it will be proportional to $y_z \boldsymbol{B}$ in the leading order. 
Eq. (\ref{dm}) implies thus that $\Delta \langle M \rangle$ should be proportional to $y_z^2\langle(d\Delta \boldsymbol{p}_1/dy_z-d\Delta \boldsymbol{p}_2/dy_z)^2\rangle$ at small $|y_z|$. 
Therefore we have also performed an initial study for the case $\tau_b=0.4 \, \rm fm/c$ and $a=1$
finding at small $y_z$ an additional $y_z^2$ dependence of both the mass and the width of $Z^0$. 
However the increase of $\Delta \langle M \rangle$  is about one order of magnitude smaller than the one observed in at $y_z=0$, while
the $\Delta \sigma$ can acquire an additional increase that is comparable to the one found at zero rapidity.
We will report about these further aspects in an upcoming longer paper.

\subsection{The centrality dependence of the leptonic invariant mass and width of $Z^0$ in the presence of the e.m. fields}
Finally we present the shifts of $Z^0$ leptonic invariant mass and its width induced by e.m. fields as a function of centrality. 
To do this one needs to calculate  the space extension, the strength and the time evolution of the magnetic field. 
Given the evolution of the first two with centrality should follow from the initial geometry, while the time dependence of the
magnetic field is the main quantity we aim to constraint,
we consider the case where the time evolution is independent of centrality.
This can serve as a baseline to interpret the future experimental results vs the centrality dependence.

\begin{figure}[h]
\centering
\includegraphics[width=1\linewidth]{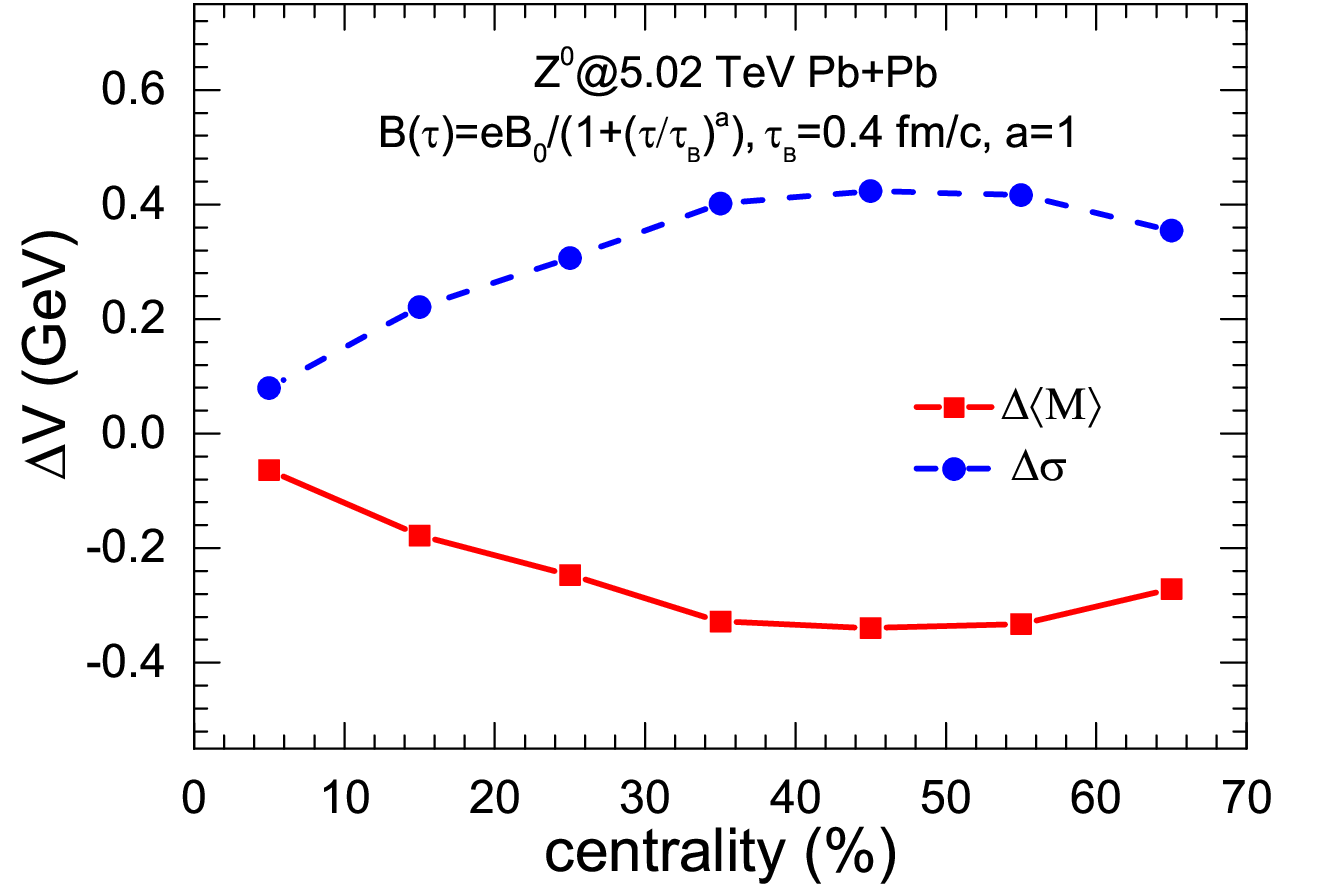}
\caption{(Color online)  The centrality dependence of $p_T$ integrated $\Delta \langle M \rangle$ and $\Delta \sigma$ of midrapidity ($|y_z|\le 0.5$) $Z^0$ boson induced by e.m. fields.}
\label{fig:centralitydependence}
\end{figure}

On the centrality dependence of the strength and the space extension of the magnetic field, we estimate it using its value in the vacuum in AA collisions at $t=0$ as well~\cite{Deng:2012pc}. The results are shown in Fig. \ref{fig:centralitydependence}, where it is seen that both $\Delta \langle M \rangle$ and $\Delta \sigma$ increase monotonically from -64 MeV to -340 MeV and from 79 MeV to 423 MeV respectively, as the centrality increases from 5\% to 45\%. 
The pattern comes as a 
balance between the increase with the impact parameter of the maximum initial value of the magnetic field and the decrease 
of the space and time extension of the fireball and of  the magnetic field as driven by the evolution of the geometry. 

At centrality around 40\% the two effects become of equal magnitude and the mass and width modifications are nearly independent of centrality.
Therefore we estimate that this is the centrality where the effects should be maximal; if experimentally the maximum is reached at smaller centrality
it would be a signature that the lifetime of the magnetic field decreases with centrality already at smaller centrality.

\section{Conclusions and Discussions}
This Letter points out a new effect that should be observable in relativistic heavy  ion collisions: the modification of both the mean value
and the width of the $Z^0$ leptonic invariant mass due to the strong initial electromagnetic field, more specifically a
decrease of the invariant mass of $Z^0$ that can be as large as few hundred MeV and the increase of the width
by a similar magnitude.
Using a wide range of reasonable parametrization for the electromagnetic field, and carrying out 
a comprehensive study of the modification of the invariant mass of $Z^0$, we find that 
 the decrease of the invariant mass $\langle \Delta M_{Z^0} \rangle$ is proportional to $\left( \int_{\tau_0}^{\tau_1}d\tau eB(\tau)\right)^n $ 
 with $n$ that has a very weak dependence on the specific behavior of the $B_y(\tau)$ and has a range of $n_M=2.16 \pm 0.16$. 
 Even more remarkable is that   the increase in the width $\Delta \sigma_{Z^0}=k_\sigma \left( \int_{\tau_0}^{\tau_1}d\tau eB(\tau)\right)^2  $
 with $k_\sigma= 6.44 \times 10^{-3}$ for all the configurations explored. 
Moreover, the shifts of both the invariant mass and the width of reconstructed $Z^0$ boson are expected to depend 
also on the rapidity of $Z^0$ quadratically and are expected to further increase the width of the invariant mass distribution. 
These modifications on the invariant mass distribution of reconstructed  $Z^0$ boson due to electromagnetic fields provide a clear probe of electromagnetic fields, which can be tested by experiments at LHC.
The main effect pointed out is novel and quite relevant in itself considering that a modification of the invariant mass of the $Z^0$
in AA collisions has never been
pointed out before, it appears to be a powerful tool to have a measure of the time integral of the magnetic field produced in relativistic
heavy  ion collisions. In the future it could be also complemented by the recent suggestions to
measure the splitting of the directed flow of $D^0$ and $D^0$ and $l^\pm$ \cite{Das:2016cwd,Sun:2020wkg,Sun:2021psy},
that instead is found to be proportional to $\tau_0 B_y(\tau_0) - \tau_1 B_y(\tau_1)$.
The scope of such studies is even more wide because a determination of the e.m. field can trigger a breakthrough in the ongoing
search for the CME, CMW and  CVE effects ~\cite{Kharzeev:2007jp,Fukushima:2008xe,Kharzeev:2009fn,Jiang:2016wve,Shi:2017cpu,Sun:2018idn,PhysRevLett.125.242301} as well as on the splitting 
of the $\Lambda$ polarization \cite{Becattini:2016gvu,Han:2017hdi,Guo:2019joy}.

\section*{ACKNOWLEDGEMENTS}
The authors acknowledge the support of INFN-SIM national project and linea di intervento 2 for HQCDyn at DFA-Unict. XNW is supported by DOE under Contract No. DE-AC02-05CH11231, by NSF under Grant No. ACI-1550228 within the JETSCAPE and No. OAC-2004571 within the X-SCAPE Collaboration.

\end{document}